\documentclass[aps,prd,superscriptaddress,nofootinbib,amsmath,amsfonts,groupedaddress]{revtex4-2}

\usepackage{setspace}
\usepackage{graphicx}
\usepackage{subcaption}
\usepackage{epsfig}
\usepackage{bm}
\usepackage{amssymb}
\usepackage{float}
\usepackage{amsmath}
\usepackage{palatino}
\usepackage{dcolumn}
\usepackage[usenames,dvipsnames]{color}
\usepackage{enumitem}
\usepackage[mathscr]{eucal}
\usepackage{hyperref}
\hypersetup{colorlinks=true,linkcolor=blue,citecolor=red,urlcolor=blue}
\def\doi{http://doi.org}

\newcommand{\be}{\begin{equation}}
\newcommand{\ee}{\end{equation}}
\newcommand{\beano}{\begin{eqnarray*}}
\newcommand{\eeano}{\end{eqnarray*}}
\newcommand{\ba}{\begin{eqnarray}}
\newcommand{\ea}{\end{eqnarray}}

\def\chariteratehelpA#1 #2\relax{%
  \chariteratehelpB#1\relax\relax%
  \ifx\relax#2\else\rlap{\charop{~}}\ \chariteratehelpA#2\relax\fi
}
\def\chariteratehelpB#1#2\relax{%
  \charop{#1}%
  \ifx\relax#2\else
    \chariteratehelpB#2\relax%
  \fi
}
\def\charop#1{\def\stacktype{L}\def\useanchorwidth{T}%
  \stackon[0pt]{#1}{\scalebox{.85}[1]{\color{red}$\sim$}}}

\usepackage{xcolor, soul}
\sethlcolor{yellow}

\begin{document}

\title{ Evolution of the universe with quintessence model in Rastall gravity}
\author{J.~K. Singh}
\email{jksingh@nsut.ac.in}
\affiliation{Department of Mathematics, Netaji Subhas University of Technology, New Delhi-110 078, India}
\author{Akanksha Singh}
\email{akanksha.ma19@nsut.ac.in}
\affiliation{Department of Mathematics, Netaji Subhas University of Technology, New Delhi-110 078, India}
\author{Shaily} 
\email{shailytyagi.iitkgp@gmail.com \text{(Corresponding author)}}
\affiliation{School of Computer Science Engineering and Technology, Bennett University, Greater Noida, India}
\affiliation{Department of Mathematics, Netaji Subhas University of Technology, New Delhi-110 078, India}

\author{Sushant G. Ghosh} 
\email{sghosh2@jmi.ac.in}
\affiliation{Centre for Theoretical Physics, Jamia Millia Islamia, New Delhi 110025, India}
\affiliation{Astrophysics Research Centre, School of Mathematics, Statistics and Computer Science, University of KwaZulu-Natal, Private Bag X54001, Durban 4000, South Africa}

\author{Sunil D. Maharaj}
\email{maharaj@ukzn.ac.za}
\affiliation{Astrophysics Research Centre, School of Mathematics, Statistics and Computer Science, University of KwaZulu-Natal, Private Bag 54001,
Durban 4000, South Africa}

\begin{abstract}
 We investigate the universe's evolution within the framework of Rastall gravity, which is an extension of the standard $\Lambda$CDM model. Utilizing a linear parametrization of the Equation of State (EoS) in a Friedmann-Lema\^{\i}tre-Robertson-Walker (FLRW) background, we constrain the model parameters through analysis of cosmic chronometers (CC),  Pantheon, Gold, Gamma Ray Burst (GRB), and Baryon Acoustic Oscillations (BAO) datasets, as well as their joint analysis, under $1\sigma$ and $2\sigma$ confidence levels, considering the Rastall parameter $\lambda$.  The constrained parameters are then used to compare our model with the standard $\Lambda$CDM model. Our findings include a detailed examination of the model's physical interpretations and demonstrate the potential for an accelerating universe expansion in later times, aligning with the observed behavior of dark energy.

\end{abstract}

\maketitle


\section{ Introduction}\label{Intro section}
The universe's accelerated expansion is one of modern cosmology's most profound and intriguing phenomena. Its discovery in the late 1990s through observation of Type Ia Supernovae \cite{SupernovaSearchTeam:1998fmf} marked a paradigm shift in our comprehension of the universe's evolution and composition. This groundbreaking finding has been corroborated by numerous independent observations, including measurements of cosmic microwave background radiation and large-scale structure surveys \cite{WMAP:2003elm, WMAP:2008lyn, SNLS:2005qlf}.

General relativity (GR) is among modern physics's most influential and robust theories. It rests on several foundational concepts and principles, including the equivalence principle, the principle of general covariance, and the Einstein field equations (EFEs). GR has undergone numerous rigorous experimental and observational tests, such as the classic verification of the bending of light by the Sun, the gravitational redshift, the precession of Mercury's orbit, and the detection of gravitational waves. Despite its successes, general relativity faces significant challenges and limitations. These include the quest for a consistent theory of quantum gravity, potential violations of the equivalence principle, and the need for a deeper understanding of the behavior of black holes and singularities.

The equivalence principle is a fundamental principle of Einstein's theory of GR. However, violations of this principle can sometimes arise, leading to gravity modifications that depart from GR predictions. In modified theories of gravity, an energy-momentum source in GR and its modifications have taken place using the divergence-free tensor, which owns a minimal coupling to the spacetime geometry \cite{Faraoni:2010pgm, Clifton:2011jh}. It is countered, though, that the attribute of the energy-momentum tensor (EMT), which results in the energy-momentum conservation law ($ \nabla_\nu T^{\mu\nu} = 0 $), is violated by the process of particle creation \cite{Parker:1971pt, Gibbons:1977mu, Birrell:1982ix, Ford:1986sy}. As a result, it is acceptable to overlook the criterion of EMT conservation and seek a new theory of gravitation. In the current era, the approach of using extended gravity theories \cite{Corda:2009re, Nojiri:2010wj, Singh:2016eom, Capozziello:2011et, Will:2014kxa, Nojiri:2017ncd} to address the critical issues with the standard cosmological model, such as dark energy \cite{SupernovaSearchTeam:1998fmf, SupernovaSearchTeam:2001qse} and dark matter \cite{Navarro:1996gj, Clowe:2006eq}, is thought to be very beneficial and appealing. It is important to note that all possible alternatives to GR should be valid. Various authors have discussed the modified theories of gravity broadly \cite{Singh:2019fpr, Singh:2022jue, Shrivastava:2021hsu, Singh:2022nfm, Singh:2018xjv, Shabani:2016dhj, Singh:2014kca, Singh:2022eun, Singh:2024ckh, Singh:2024gtz, Shaily:2024nmy, Goswami:2023knh, Shaily:2024rjq, Singh:2024kez, Singh:2022wwa, Singh:2023gxd, Singh:2022ptu, Singh:2022gln, Goswami:2022vfq, Nagpal:2018mpv}.

One such modification is Rastall gravity, which Peter Rastall put forth in 1972 \cite{Rastall:1972swe}. Rastall's work draws attention to the fact that the EMT conservation law is examined under specific conditions, like within flat Minkowski spacetime or, in particular, in weak gravitational fields, and this is why there are doubts about its applicability in curved spacetime. In light of Rastall's claim, the assumption that the covariant derivative of the EMT vanishes, no longer holds. Instead, we consider a vector field that aligns with the gradient of the Ricci scalar, expressed as $ \nabla_\nu T^{\mu\nu} \propto \nabla^\mu R $. Since the curvature in Rastall's theory is associated with a violation of the laws of energy conservation and momentum, it can be considered a classical formulation of quantum phenomena \cite{Batista:2011nu}. There is also the curvature-matter theory of gravity, which, as explored in works \cite{Nojiri:2004bi, Koivisto:2005yk, Bertolami:2007gv, Harko:2014gwa}, introduces a non-minimal coupling between matter and geometry, leading to the breakdown of the conventional energy-momentum conservation law.

One major limitation of Rastall's theory is its need for a Lagrangian formulation. Numerous studies have endeavored to establish a Lagrangian for Rastall theory \cite{DeMoraes:2019mef, Shabani:2020wja, Fabris:2020uey, Santos:2017nxm}. Efforts have been made to derive the Lagrangian for Rastall gravity from an action principle using Weyl's geometry \cite{Smalley:1993ab}. Shabani and Ziaie tackled this issue by considering the matter content as a perfect fluid described by a barotropic equation of state \cite{Shabani:2020wja}. They confirmed that the Rastall theory can be associated with the $f(R, T)$ gravity Lagrangian. Notably, it has been shown that Rastall theory is a particular case within the more extensive framework of $f(R, T)$ gravity \cite{Santos:2017nxm}. Furthermore, Darabi et al. \cite{Darabi:2017coc} provided a clear distinction between Rastall gravity and general relativity, countering Visser's assertion that the two are equivalent \cite{Visser:2017gpz}.

Over the past several years, substantial progress has been made in exploring various facets of Rastall gravity. This includes implications for the structure of neutron stars \cite{Oliveira:2015lka, Maurya:2020}, the influence of the Rastall parameter on dark energy perturbations \cite{Ziaie:2019gan}, and its integration with the Brans-Dicke scalar field \cite{Carames:2014twa}. The theory has also been applied in the context of helium nucleosynthesis \cite{Al-Rawaf:2005dpt}. Additionally, researchers have investigated different cosmic epochs within this framework \cite{Al-Rawaf:1994muv, Majernik:2002gd, Fabris:2012hw, Singh:2020akk}, providing a comprehensive analysis of both theoretical and observational aspects of Rastall gravity \cite{Akarsu:2020yqa, Singh:2022ocv}. Notably, Rastall gravity is free from entropy and age problems \cite{Fabris:1998hr}, and offers explanations for both inflationary and accelerated expansion phases \cite{Fabris:2011wz, Moradpour:2017shy}. According to some authors \cite{Moradpour:2017shy}, a non-minimal coupling between pressure-less matter and geometry in Rastall gravity could emulate dark energy behavior, potentially accounting for the current phase of cosmic acceleration. The Rastall gravity model has been thoroughly examined across the cosmic timeline, from early inflation to accelerated expansion, including the matter-dominated era \cite{Das:2018dzp}. Furthermore, the Rastall cosmological model proposed in \cite{Khyllep:2019odd} aligns with the $\Lambda$CDM model in the late universe. A suitable Rastall parameter has also been suggested as a possible solution to the initial singularity problem \cite{Das:2018dzp}. Rastall gravity is consistent with gravitational lensing phenomena \cite{Abdel-Rahman:2001lmc, Abdel-Rahman:2005evy} and has been extensively studied in various cosmological contexts \cite{Al-Rawaf:1995xkt, Lindblom:1982, Batista:2010nq, Batista:2012hv, Silva:2012gn, Santos:2014ewa, Oliveira:2016ooo, Moradpour:2015ymo, Moradpour:2016fur, Yuan:2016pkz, Heydarzade:2017wxu, Heydarzade:2016zof, Moradpour:2016ubd, Licata:2017rfx, Darabi:2017tay, Hansraj:2018zwl, Halder:2019akt, Lin:2018dgx, Cruz:2019jiq, Ziaie:2020ord, Ghaffari:2020nnk, Yu:2019cku, Gogoi:2021dkr, Shaily:2024xho}. This extensive body of work highlights the versatility and potential of Rastall gravity in addressing some of the most profound questions in cosmology.

Our work is structured as follows: Section \ref{Field equations section} focuses on the theoretical framework of the model. In Section \ref{Obs}, we present contour plots for the Hubble dataset consisting of $ 77 $ points, the Pantheon dataset of $ 1048 $ points, the Gold dataset having $ 185 $ points, the Gamma Ray Burst comprising of $ 162 $ points, and the BAO data for $ 6 $ points to constrain the model parameters $\omega_0$, $\omega_1$, and $\lambda$. This section also includes an analysis of error bar plots using various observational datasets. Section \ref{Fourth section} provides graphical representations of all cosmological parameters about redshift $z$. For a more comprehensive analysis, Section \ref{Fifth section} examines the energy conditions, the statefinder diagnostic, and the properties of the scalar field $\phi$. In Section \ref{Perturbations section}, we employ linear perturbation theory to investigate the stability of the model. Finally, Section \ref{Conclusions section} summarizes our findings and concluding remarks.

\section{ Framework of the Model}\label{Field equations section}
We aim to derive solutions by meticulously accounting for the gravitational background and the matter source. To investigate the Rastall gravity framework, we incorporate the Rastall field equations alongside the matter field equations. In this modification of GR, the covariant conservation equation of the energy-momentum tensor represented as $ \nabla_\nu T^{\mu \nu} = 0 $ can be written in the more generalized form as \cite{Sekhmani:2024jli}
\begin{equation}\label{1}
\nabla_{\nu} T^{\mu \nu} = u^{\mu}.
\end{equation}
To convert this again in GR, the right side of Eq. (\ref{1}) must be zero. Therefore, the four-vector $u^{\mu}$ can be taken as,
\begin{equation}\label{2}
u^{\mu}=\lambda \nabla^\mu R,
\end{equation}
where $ \lambda $ is a free parameter called Rastall’s parameter.  For $ \lambda=0 $, the theory reduces to GR, while non-zero values of $ \lambda $ introduce new dynamics. In Rastall gravity, $ \lambda $ represents the deviation from the standard conservation of EMT. The Rastall parameter $ \lambda $ modifies the EFEs, particularly affecting the Hubble parameter, matter density, and isotropic pressure. Several studies have attempted to constrain $ \lambda $ using observational data \cite{Akarsu:2020yqa, Astorga-Moreno:2024okq, Li:2019jkv, Bishi:2023mwv, Afshar:2023uyw, Tang:2019dsk, Waseem:2024nba}. Its expected range is often small, typically constrained around values close to zero but non-zero with certain error margins. The estimated range for values of $ \lambda $ often varies depending on the specific dataset or cosmological model being used. From the above two equations, one can obtain the Rastall field equations as \cite{Rastall:1972swe, Capone:2009xm}
\begin{equation}\label{3}
R_{\mu \nu}-\frac{1}{2}\left(1-2 K_R \lambda \right)Rg_{\mu \nu} = K_G T_{\mu \nu},  
\end{equation}
where $ g_{ \mu\nu} $ is the metric tensor, $ R_{\mu \nu} $ denotes the Ricci tensor, and $ K_R $ and $ K_G $ denote the Rastall gravitational constant (RGC) and the Einstein gravitational constant (EGC), respectively. Also, the trace of the Eq. (\ref{3}) yields
\begin{equation}\label{4}
R=\frac{\kappa}{4 K_R \lambda-1}T,~\lambda \neq \frac{1}{4 K_R}.
\end{equation}
The field equation with the cosmological constant $ \Lambda $ is
\begin{equation}\label{5}
G_{\mu \nu} + \Lambda g_{\mu \nu} + K_R \lambda g_{\mu \nu} R = K_G T_{ \mu \nu},
\end{equation}
where $ G_{\mu \nu} $ stands for the standard Einstein tensor.

Now, we consider the FLRW line element
	\begin{equation}\label{6}
		ds^{2} = -dt^{2}+a^{2}(t) \left(\frac{dr^2}{1-k r^2}+r^2\left(d\theta^2+\sin^2\theta d\phi^2\right)\right),
	\end{equation}
where $ a(t) $ is used for the scale factor, and $ k $ is the parameter that takes the values $ 0, +1, -1 $ for the flat space, closed space, and open space sections, respectively.

In Eq. (\ref{3}), we consider both parameters $ K_R $ and $ K_G $ equal to $ K $, treated as the free parameters essential for the model's requirements. This treatment is done with careful consideration of maintaining the compensation of the Bianchi identity $ {G^{\mu\nu}}_{;\nu} = 0 $ and adhering to the constraint $ {T^{\mu \nu}}_{;\nu} = (\lambda R)^{;\mu} $ within Rastall gravity without loss of generality. Here, the parameter $ \lambda $ is a non-zero constant referred to as the Rastall parameter. It's noteworthy that setting $ \lambda = 0 $ results in recovering Einstein's equations within the General Theory of Relativity (GTR) whenever $ {T^{\mu \nu}}_{;\nu} = 0 $ \cite{Rastall:1972swe, Capone:2009xm, Singh:2020akk}. To maintain consistency and avoid potential issues, we exclude the cases where $ K\lambda = \frac{1}{4} $ and $ K\lambda = \frac{1}{6} $, as these values may lead to undefined quantities \cite{Rastall:1972swe, Capone:2009xm, Singh:2020akk, Das:2018dzp}.
In Rastall gravity, the field equations for the perfect fluid tensor $ T^{\mu}_{\nu} =$ diag $(-\rho,p,p,p) $ are given by	
	\begin{equation}\label{7}
		3(1-4K\lambda)H^2-6K\lambda\dot{H}+3(1-2K\lambda)\frac{k}{a^2} = K\rho,
	\end{equation}
	\begin{equation}\label{8}
		3(1-4K\lambda)H^2+2(1-3K\lambda)\dot{H}+(1-6K\lambda)\frac{k}{a^2} = -Kp,
	\end{equation}
where $ \rho $ and $p$ are the energy density and isotropic pressure, respectively. The Hubble parameter is related to the scale factor as $ H=\frac{\dot{a}}{a} $, where $\dot{a}$ denotes the derivative of $a$ with respect to cosmic time $ t $. 

Also, the energy density and isotropic pressure are correlated by the equation of state (EoS) as
	\begin{equation}\label{9}
		p = \omega\rho.
	\end{equation}
	
Most of the observational datasets describe results as functions of redshift $ z $ instead of cosmic time $ t $. To ensure consistency with observational data, we will use the relation $$ \dot{H}(t) = -\left(1+z\right)H(z)\frac{dH}{dz} $$ to write the Friedmann equations (\ref{7}), (\ref{8}) in terms of redshift $ z $ as follows
	\begin{equation}\label{10}
	3(1-4K\lambda)H^2+6K\lambda (1+z)H\frac{dH}{dz}+3(1-2K\lambda)k(1+z)^2 = K\rho,
	\end{equation}
	\begin{equation}\label{11}
	3(1-4K\lambda)H^2-2(1-3K\lambda)(1+z)H\frac{dH}{dz}+(1-6K\lambda)k(1+z)^2 = -Kp,
	\end{equation}
where $ H=H(z) $, $ \rho=\rho(z) $, and $ p=p(z) $. 
We normalize the coupling constant ($ K $) to reduce the mathematical complexities by taking $ K = 1 $.
		

Using Eqs. (\ref{9}), (\ref{10}) in (\ref{11}) with $ K = 1 $ and for a flat universe ($ k=0 $) yields
\begin{equation}\label{12}
    3(1-4\lambda)[1+\omega(z)]H^2(z) - \left[2-6\lambda(1+\omega(z))\right] \left(1+z\right)H(z)\frac{dH}{dz} = 0.
\end{equation}

The above equation has two unknown functions, $ \omega $ and $ H $. We need to parameterize one of these unknown functions to get the solution and proceed with our analysis. With the help of numerous studies \cite{Weller:2001gf, Linder:2002et, Zhao:2005vj, Nesseris:2005ur, Goliath:2001af}, we consider the linear parametrization of the EoS parameter $ \omega (z)$ as
\begin{equation}\label{13}
\omega (z) = \omega_0 + \omega_1 z,
\end{equation}
where $ \omega_0 $ and $ \omega_1 $ are arbitrary constants. This parameterization for the dark energy equation of state is motivated by theoretical considerations, observational data, and the ease of working with this form. Since this parametrization has few parameters, it will help us study the various cosmological behaviors without adding unnecessary complexity.

Using Eq. (\ref{13}) in Eq. (\ref{12}), we get
\begin{equation}\label{14}
3(1-4\lambda)(1+\omega_0+\omega_1z)H^2 - \left(2-6\lambda(1+\omega_0+\omega_1z)\right)(1+z)H\frac{dH}{dz} = 0.
\end{equation}
The Hubble parameter is obtained by solving Eq. (\ref{14}) as
\begin{equation}\label{15}
H(z) = \exp(B),
\end{equation} 
where
\begin{equation}\label{15a}
B = \frac{(-1+4\lambda) \left( 3\lambda \ln(1+z) - \ln(1-3\lambda(1+\omega_0+\omega_1z)) + 3\lambda \ln(1+z) (\omega_0-\omega_1) \right)}{2\lambda(-1+3\lambda+3\lambda\omega_0-3\lambda\omega_1)}. \nonumber
\end{equation}

The deceleration parameter $ q $ can also be obtained in terms of $ z $ as
\begin{equation}\label{16}
q(z) = \frac{-1+6\lambda+(6\lambda-3)\omega_0+3(2\lambda-1)z\omega_1}{-2+6\lambda+6\lambda(\omega_0+\omega_1z)}.
\end{equation}
Using Eq. (\ref{15}) in Eq. (\ref{10}), we get
\begin{equation}\label{17}
\rho(z) = \frac{3(-1+4\lambda) \exp(2B)}{-1+3\lambda(1+\omega_0+\omega_1z)}.
\end{equation}
From Eqs. (\ref{9}) and (\ref{17}), we get
\begin{equation}\label{18}
p(z) = \frac{3(-1+4\lambda)(\omega_0+\omega_1z) \exp(2B)}{-1+3\lambda(1+\omega_0+\omega_1z)}.
\end{equation}
	
\section{Observational Constraints}\label{Obs}
\qquad In this section, we utilized the observational cosmic chronometers dataset (CC), Pantheon, Gamma Ray Burst, Gold dataset, Baryonic Acoustic Oscillations (BAO), and their combined dataset to derive the optimal values for the model parameters $ \omega_0 $, $ \omega_1 $, and $ \lambda $. And, for this purpose, we use the MCMC (Markov Chain Monte Carlo) technique in the emcee Python environment for each dataset \cite{Foreman-Mackey:2012any}. The resultant optimal values for these model parameters are detailed in Table \ref{tab1}.

\begin{figure}
    \begin{center}
        \subfloat[]{\label{Hz} \includegraphics[scale=0.62]{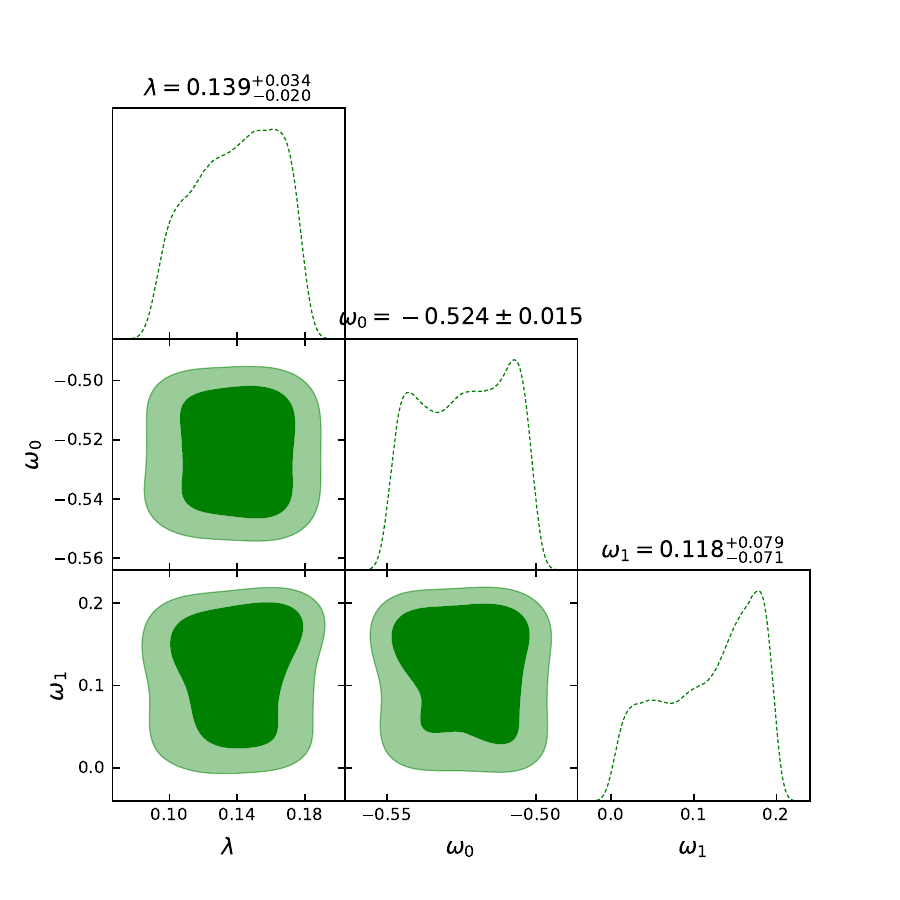}}
        \subfloat[]{\label{PGG} \includegraphics[scale=0.62]{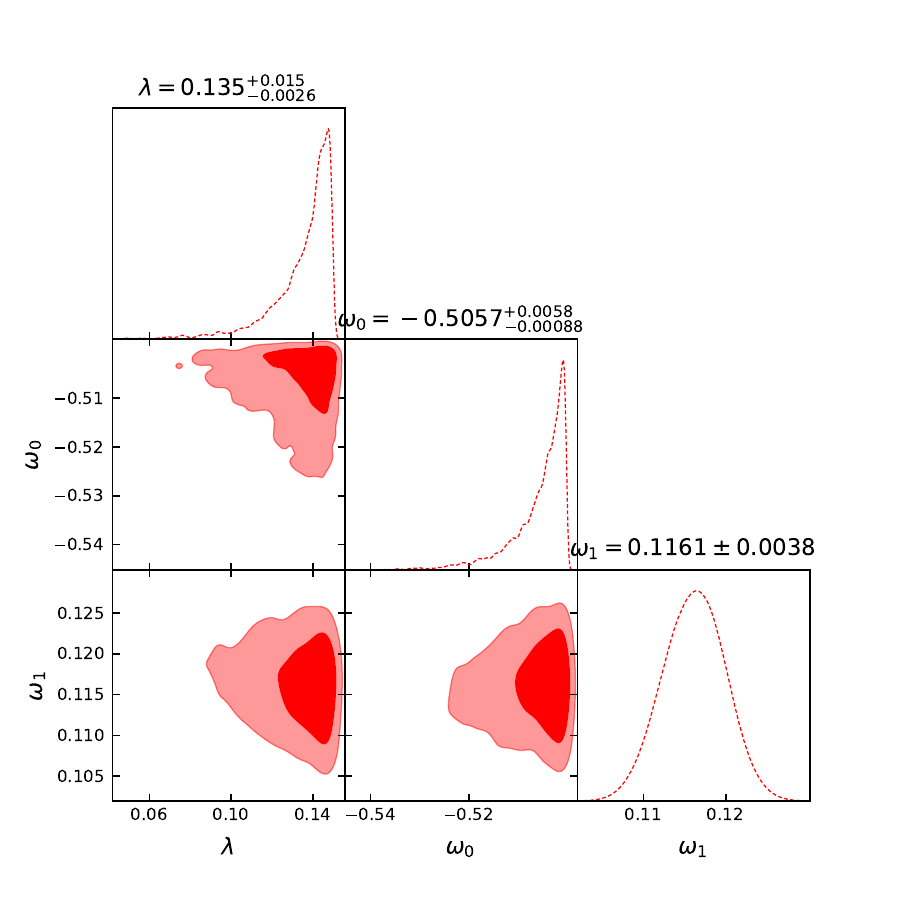}}\\
         \subfloat[]{\label{HPGG} \includegraphics[scale=0.62]{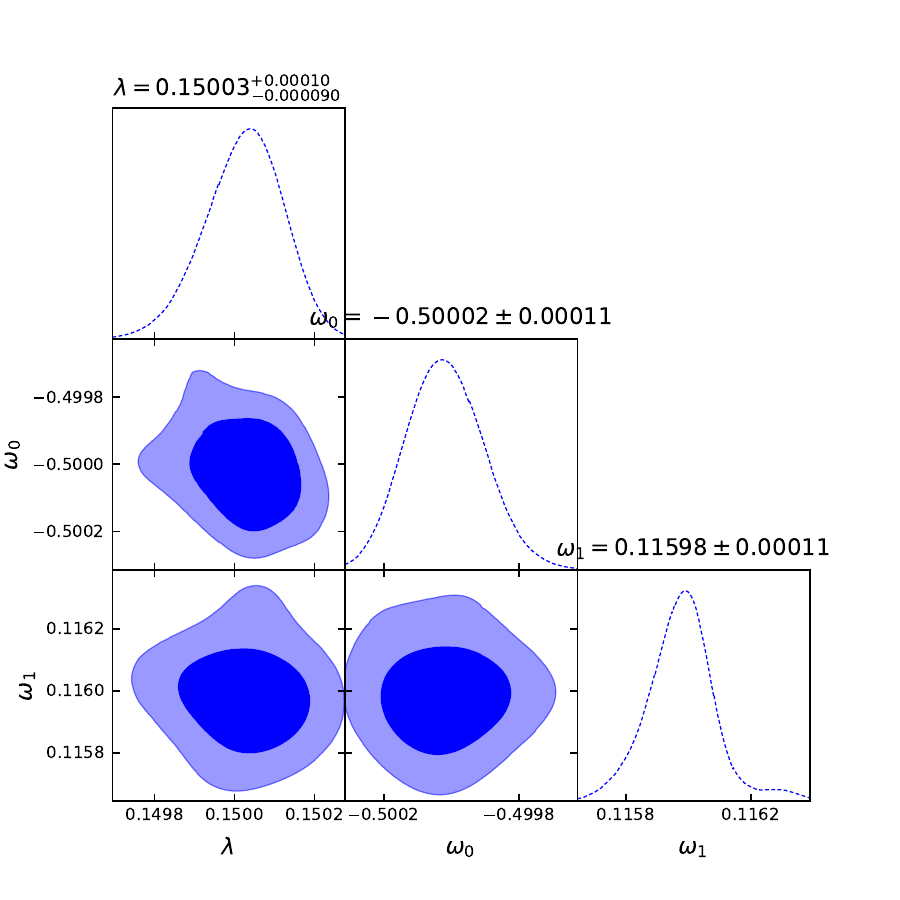}}
        \subfloat[]{\label{HPGGB} \includegraphics[scale=0.62]{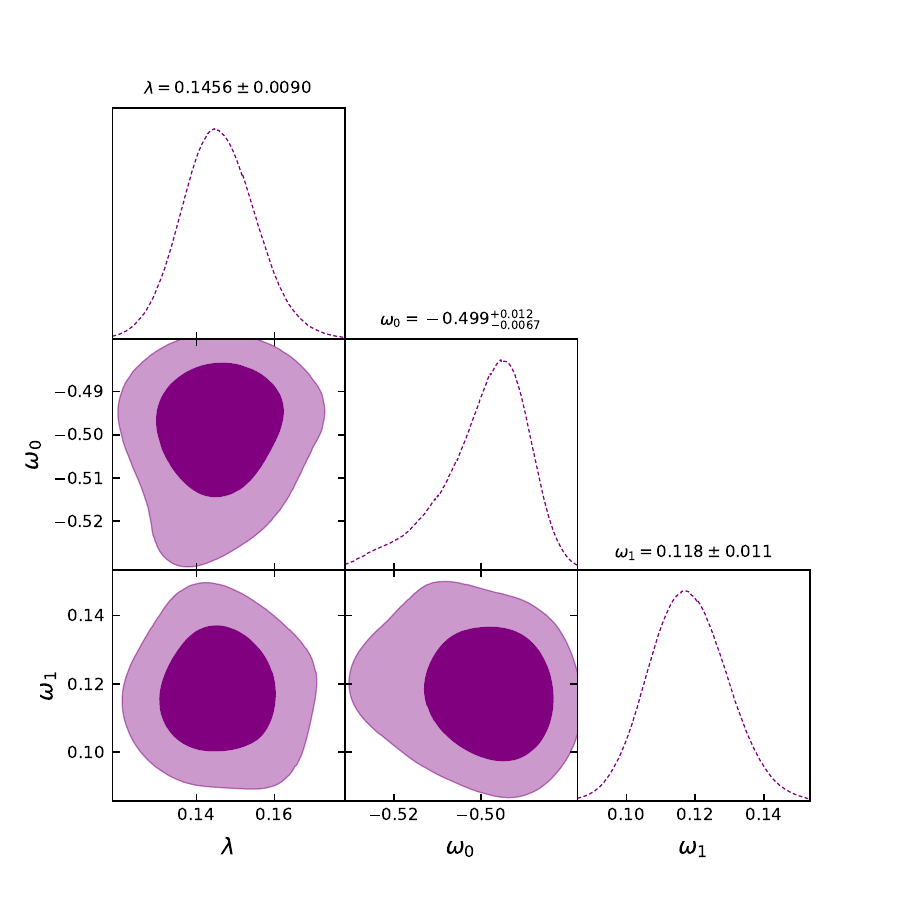}}
    \end{center}
\caption{ {\it The Likelihood contours for CC, PGG (Pantheon+Gold+Gamma Ray Burst), CC $+$ PGG, and joint analysis of CC $+$ PGG $+$ BAO with $ 1\sigma $ and $ 2\sigma $ confidence levels. }}
\label{hz2}
\end{figure}

\subsection{Cosmic Chronometers Dataset}
\qquad In previous investigations, a cosmic chronometer is a galaxy that evolves passively without ongoing star formation. The Differential Age (DA) method is the tool to gather such data, which remains independent of specific cosmological models. Consequently, this data becomes invaluable for scrutinizing various cosmological models. Our current study uses Cosmic Chronometer (CC) data comprising 77 data points collated from references \cite{Shaily:2022enj, Singh:2023ryd, Singh:2024zvm, Singh:2024aml}. Employing a chi-square minimization approach, we determine the optimal values for $ \omega_0 $, $ \omega_1 $, and $\lambda $ as follows:

\begin{equation}\label{19}
\chi _{HD}^{2}(\omega_0,\omega_1,\lambda)=\sum\limits_{i=1}^{77} \frac{[H_{th}(\omega_0,\omega_1,\lambda,z_{i})-H_{obs}(z_{i})]^2}{\sigma _{z_i}^2},
\end{equation}
where $ H_{th}(\omega_0,\omega_1,\lambda,z_{i}) $ and $ H_{obs}$ indicate the theoretical and observed values, respectively, where $ \omega_0,~\omega_1, ~\lambda $ are the model parameters, and $ \sigma_{(z_{i})} $ shows the standard deviation for each $ H(z_i) $.

\begin{figure}
\centering  
	 { \includegraphics[scale=0.472]{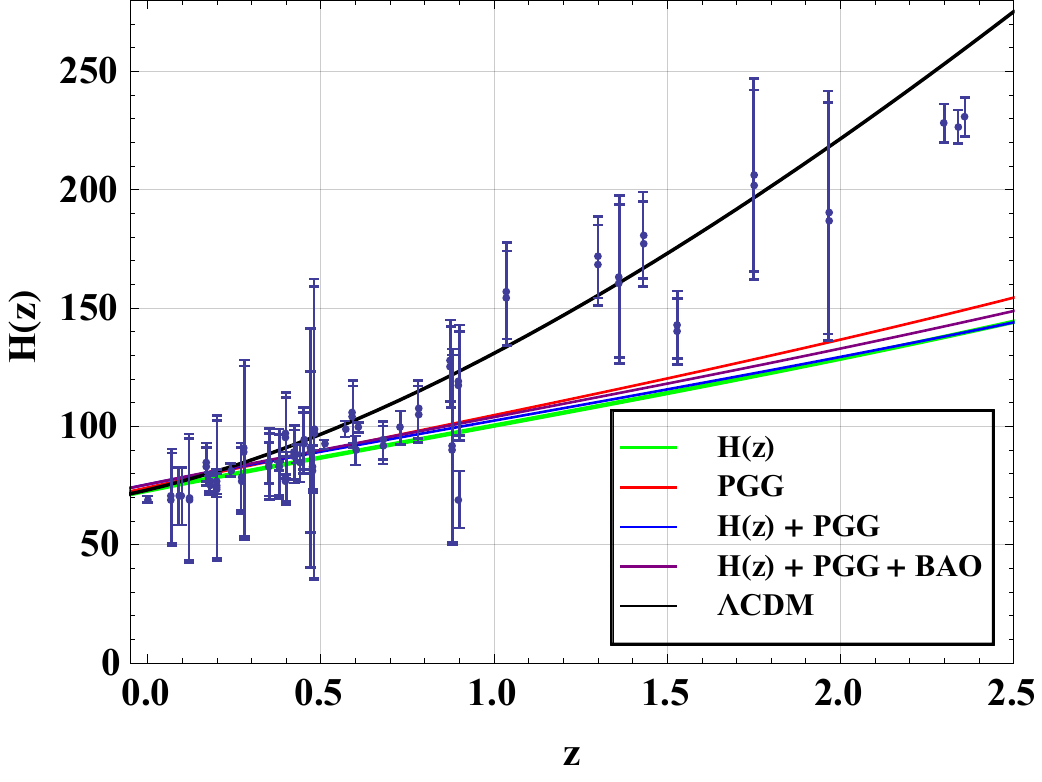}}\hfill
	 { \includegraphics[scale=0.455]{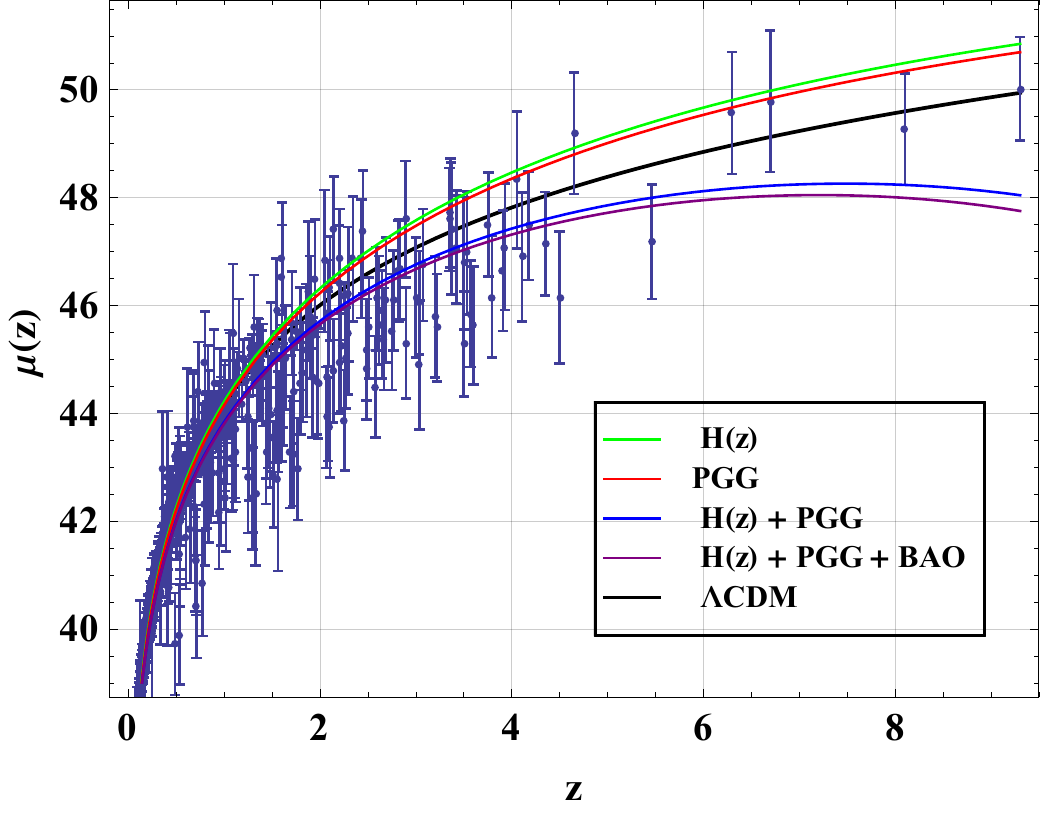}}
\caption{ {\it The error bar plots of the Hubble parameter $ H(z) $ (left) and the distance modulus $ \mu(z) $ (right) concerning redshift $ z $ for CC, PGG (Pantheon+Gold+Gamma Ray Burst), CC $+$ PGG, and for the joint analysis of CC, PGG, and BAO data in the presence of the Rastall parameter $ \lambda $. These plots indicate the degree of consistency of our model with $ \Lambda $CDM.}}
\label{err-z}
\end{figure}

\subsection{Pantheon, Gamma Ray Burst and Gold Datasets}
\qquad This model uses the PGG compilation dataset with $ 1395 $ points for the redshift range $ 0<z<9.3 $ \cite{SupernovaCosmologyProject:2011ycw}. This data consists of $ 1048 $ data points of Pantheon data, which is collected from various supernovae surveys for the redshift range $ 0.01 \leq z \leq 2.26 $, $ 162 $ data points of Gamma Ray Burst for $ 0.03 \leq z \leq 9.3 $, and $ 185 $ data points of Gold data for the redshift range $0.01 \leq z \leq 1.755 $ \cite{Pan-STARRS1:2017jku, Riess:1998dv, Jha:2005jg, Hicken:2009df, Contreras:2009nt, SDSS:2014irn, Izzo:2015vya, Demianski:2016dsa, Amati:2018tso}. These datasets play an important role in investigating the expansion rate of the universe. With these data points, the minimum value of the chi-square function $ \chi_{PGG}^{2} $ can be written as
\begin{equation}\label{20}
\chi _{PGG}^{2}( \omega_0,\omega_1,\lambda)=\sum\limits_{i=1}^{1395}\left[ \frac{\mu_{th}(\omega_0,\omega_1,\lambda,z_{i})-\mu_{obs}(z_{i})}{\sigma _{\mu(z_{i})}}\right] ^2,
\end{equation}
where $ PGG $ represents the combined dataset of Pantheon, Gamma Ray Burst, and Gold data and $ \sigma_{\mu(z_{i})} $ denotes the standard error in the observed value. $ \mu_{th} $ and $ \mu_{obs} $ stand for our model's theoretical and observed distance modulus. Also, the theoretical distance modulus $ \mu(z) $ can be written in terms of the apparent magnitude $ m $ and absolute magnitude $ M $ of a standard candle as
\begin{equation}\label{21}
\mu(z)= m-M = 5~\log_{10} D_L(z)+\mu_{0},
\end{equation}
where $ D_L(z) $ is the luminosity distance and $ \mu_0 $ is the nuisance parameter for a flat universe, which are given by
\begin{equation}\label{22}
D_L(z)=(1+z) c \int_0^z \frac{1}{H(z^*)}dz^*,
\end{equation}
and
\begin{equation}\label{23}
\mu_0= 5~\log_{10}\Big(\frac{H_0^{-1}}{1 \rm Mpc}\Big)+25,
\end{equation}
respectively.

\subsection{Baryonic Acoustic Oscillations (BAO) Dataset}
\qquad BAO investigates the oscillations in the early universe caused by cosmic perturbations in a fluid consisting of photons, baryons, and dark matter, with interactions mediated through Thomson scattering. In this study, we analyze a set of BAO distance measurements from various surveys, including SDSS(R) \cite{Padmanabhan:2012hf}, the 6dF Galaxy Survey \cite{Beutler:2011hx}, BOSS CMASS \cite{BOSS:2013rlg}, and three independent measurements from the Wiggle Z survey \cite{Blake:2012pj, Blake:2011en, SDSS:2009ocz}. To perform the BAO analysis, the following equations are utilized. The distance-redshift ratio is obtained from the equation
\begin{equation}
    d_z= \frac{r_s(z^*)}{D_v(z)},
\end{equation}
where $ r_s(z^*) $ denoted the co-moving sound horizon at the time when photons decouple \cite{SDSS:2005xqv}, defined as $ r_s(a) = \int_0^a \frac{c_s da}{a^2 H(a)} $ and $ z^* $ is the photons decoupling redshift. Here we consider $ z^* = 1090 $ for the analysis. The dilation scale is denoted by $ D_v(z)$ which is denoted by 
\begin{equation}
    D_v(z) = \bigg(\frac{z {d_A}^2 (z)}{ H(z)} \bigg)^{\frac{1}{3}}, 
\end{equation}
where $ d_A(z) $ represents the angular diameter distance, which is defined as $ d_A(z) = c \int_0^{z} \frac{dz'}{H(z')} $.

The $ \chi^{2}_{BAO} $ corresponding to BAO measurements is given by \cite{Giostri:2012ek}
\begin{equation}
    \chi_{BAO}^2 = X^{T} C^{-1} X.
\end{equation}

\begin{equation}
X = 
\begin{bmatrix}
$$ \frac{d_A(z^*)}{D_v(0.106)}-30.84$$ \\
$$ \frac{d_A(z^*)}{D_v(0.35)}- 10.33 $$\\
$$ \frac{d_A(z^*)}{D_v(0.57)}- 6.72 $$\\
$$ \frac{d_A(z^*)}{D_v(0.44)}- 8.41 $$\\
$$ \frac{d_A(z^*)}{D_v(0.6)}- 6.66 $$\\
$$ \frac{d_A(z^*)}{D_v(0.73)}- 5.43 $$
\end{bmatrix}
\end{equation}

The inverse of the covariance matrix ($ C^{-1} $) is given by \cite{Giostri:2012ek}

\begin{equation}
C^{-1} = 
\begin{bmatrix}
 0.52552 & -0.03548 & -0.07733 & -0.00167 & -0.00532 & -0.0059 \\
 -0.03548 & 24.9707 & -1.25461 & -0.02704 & -0.08633 & -0.09579 \\
 -0.07733 & -1.25461 & 82.9295 & -0.05895 & -0.18819 & -0.20881 \\
 -0.00167 & -0.02704 & -0.05895 & 2.9115 & -2.98873 & 1.43206 \\
 -0.00532 & -0.08633 & -0.18819 & -2.98873 & 15.9683 & -7.70636 \\
 -0.0059 & -0.09579 & -0.20881 & 1.43206 & -7.70636 & 15.2814 \\
\end{bmatrix}
\end{equation}

\subsection{Joint Datasets}
\qquad By performing joint statistical analysis using the CC, PGG, and BAO datasets, we can obtain stronger constraints. Therefore, the chi-square function for joint datasets can be written as 
\begin{eqnarray}\label{24}
\chi_{Joint}^2= \chi_{HD}^2 + \chi_{PGG}^2 + \chi_{BAO}^2.
\end{eqnarray}

\begin{table}
	\caption{Best-fit values of the model parameters $ \omega_0 $, $ \omega_1 $, and $ \lambda $ for $ D_1 = H(z) $, $ D_2 = $ Pantheon+Gold+Gamma Ray Burst, $ D_3 = H(z) $+Pantheon+Gold+Gamma Ray Burst, and $ D_4 = H(z) $+Pantheon+Gold+Gamma Ray Burst+BAO.}
	\begin{center}
		\label{tab1}
		\begin{tabular}{l c c c r} 
			\hline\hline
			\\
			{Dataset} &  ~~~~~~~ $ \omega_0 $ & ~~~~~  $ \omega_1 $ & ~~~~~~  $ \lambda $ & ~~~~~~  $ z_{tr}$\footnote{ transition state from deceleration to acceleration} 
			\\
			\\
			\hline
			\\      
			{ $ D_1 $ }  &   ~~~~~~~ $ -0.524 \pm 0.015 $   &  ~~~~~ $ 0.118^{+0.079}_{-0.071} $ &  ~~~~~ $ 0.139^{+0.034}_{-0.020} $ &  ~~~~~ $ 3.59023 $ 
			\\
			\\
			{ $ D_2 $ }    & ~~~~~~~ $ -0.5057^{+0.00580}_{-0.00088} $   &  ~~~~~ $ 0.1161 \pm 0.0038 $ &  ~~~~~ $ 0.135^{+0.0150}_{-0.0026} $ &  ~~~~~ $ 3.60846 $    
			\\
			\\
			{ $ D_3 $ }  &  ~~~~~~~$ -0.50002 \pm 0.00011 $  & ~~~~~ $ 0.11598 \pm 0.00011 $ &  ~~~~~ $ 0.15003^{+0.00010}_{-0.00009} $ &  ~~~~~ $ 3.90138 $    
			\\
			\\
                { $ D_4 $ }  &  ~~~~~~~$ -0.499^{+0.0120}_{-0.0067} $  & ~~~~~ $ 0.118 \pm 0.011 $ &  ~~~~~ $ 0.1456 \pm 0.009 $ &  ~~~~~ $ 3.72506 $
                \\
                \\
			\hline\hline  
		\end{tabular}    
	\end{center}
\end{table}
Let's focus on the constraints related to the novel parameters $ \omega_0 $ and $ \omega_1 $, designated as the model parameters, within the context of the Rastall parameter $ \lambda $. In Fig. \ref{hz2}, we observe likelihood contours encompassing the optimal-fit values of these parameters across different confidence levels. Furthermore, Table \ref{tab1} provides a concise overview of the optimal-fit values derived for the model parameters $ \omega_0 $, $ \omega_1 $, and the Rastall parameter $ \lambda $. 

Fig. \ref{err-z} takes us further, illustrating error bar plots for the Hubble parameter $ H(z) $ and the distance modulus $ \mu(z) $. These graphs are plotted using the Mathematica software. For the left graph, error bars are drawn for $ 77 $ points of the $ H(z) $ dataset, while the right graph uses $ 1395 $ points of the PGG compilation observational dataset to draw the error bars. These plots are plotted with the help of the function $ H(z) $ mentioned in Eq. (\ref{15}) for the best-fit values of the model parameters obtained through the observational datasets CC, PGG, CC+PGG, and joint analysis (CC+PGG+BAO) in the presence of the Rastall parameter $ \lambda $. The graphs in Fig. \ref{err-z} quantify the deviations of our model from the standard $ \Lambda $CDM model. To create the curves corresponding to the $ \Lambda $CDM model and the $ \mu(z) $ function, we have employed the Hubble parameter value, $ H_0=73.04\pm1.04~{\rm km} ~{\rm s}^{-1}{\rm Mpc}^{-1} $, from the Riess et al. 2022 results \cite{Riess:2021jrx}. The trajectories in Fig. \ref{err-z} move closer to the $ \Lambda $CDM curve as one moves closer to the present epoch ($ z=0 $). The left panel shows that our model aligns closely with $ \Lambda $CDM for the redshift range $ 0 \leq z < 0.5 $, and the right panel indicates a similar trend for $ 0 \leq z < 4 $.
\begin{figure}
\centering
	 { \includegraphics[scale=0.45]{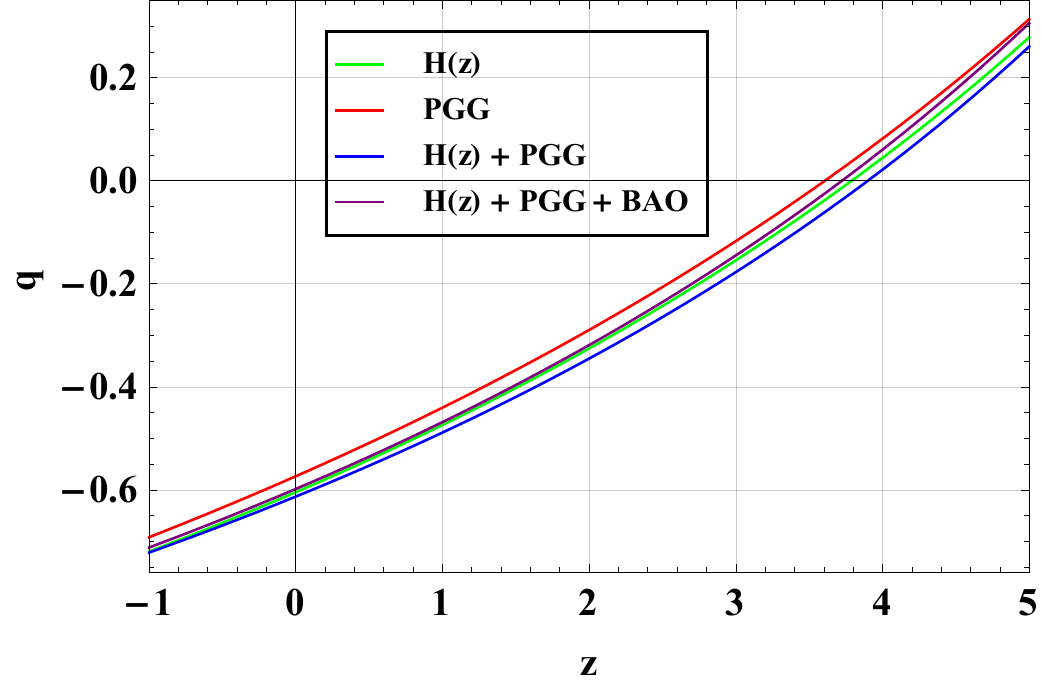}}\hfill 
	 { \includegraphics[scale=0.44]{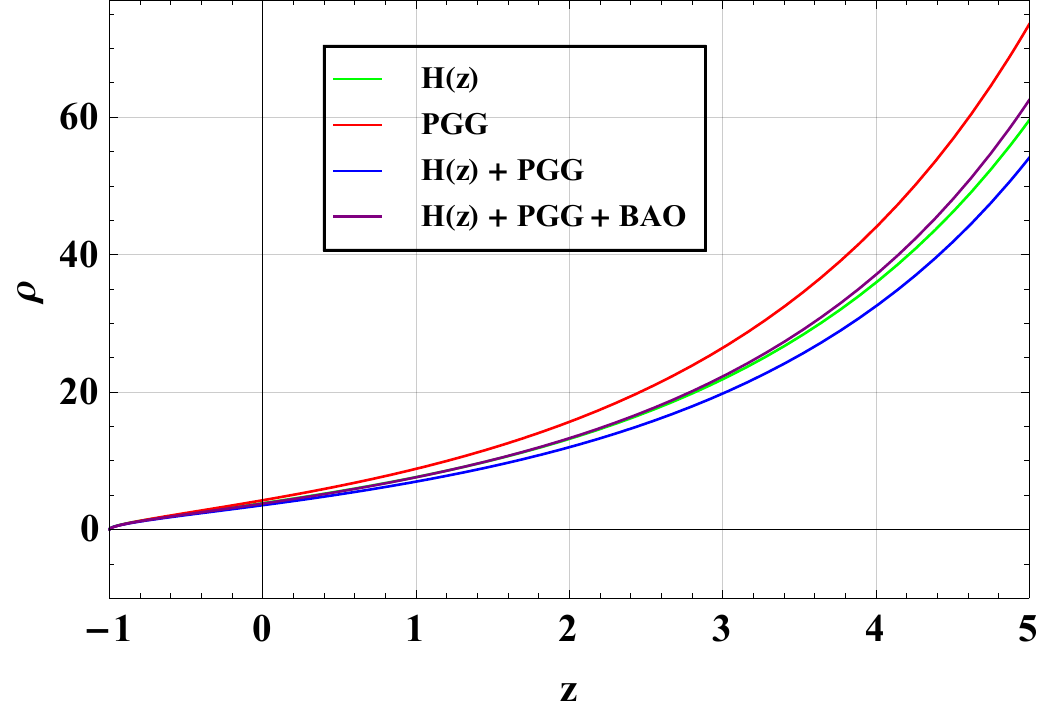}}\par  
	 { \includegraphics[scale=0.44]{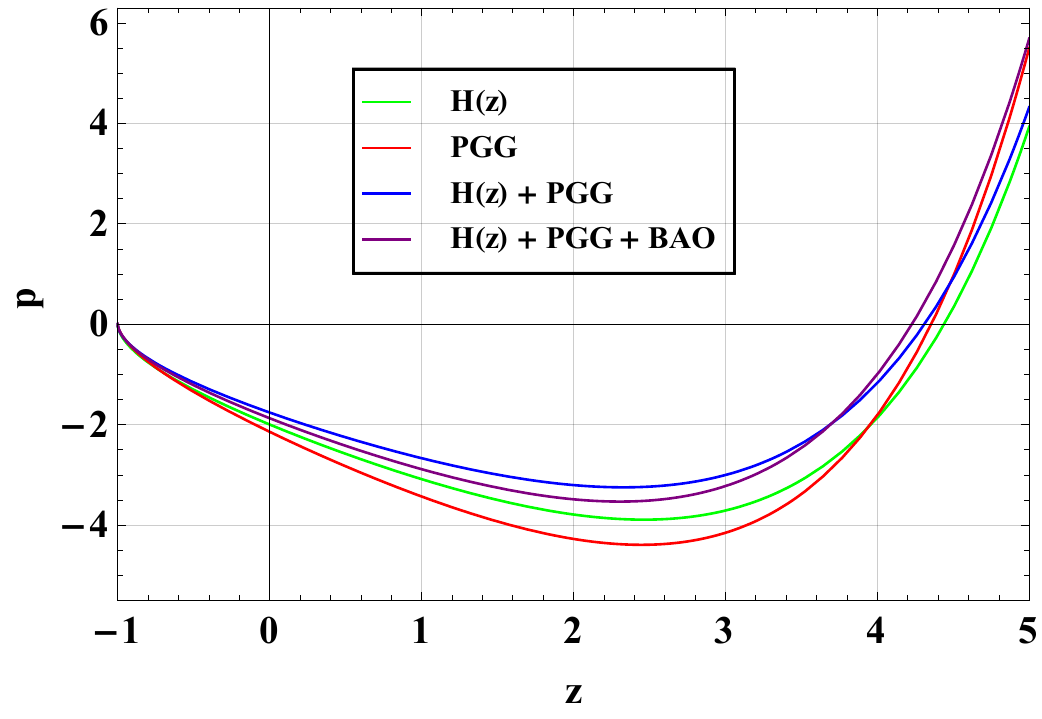}}\hfill 
	 { \includegraphics[scale=0.44]{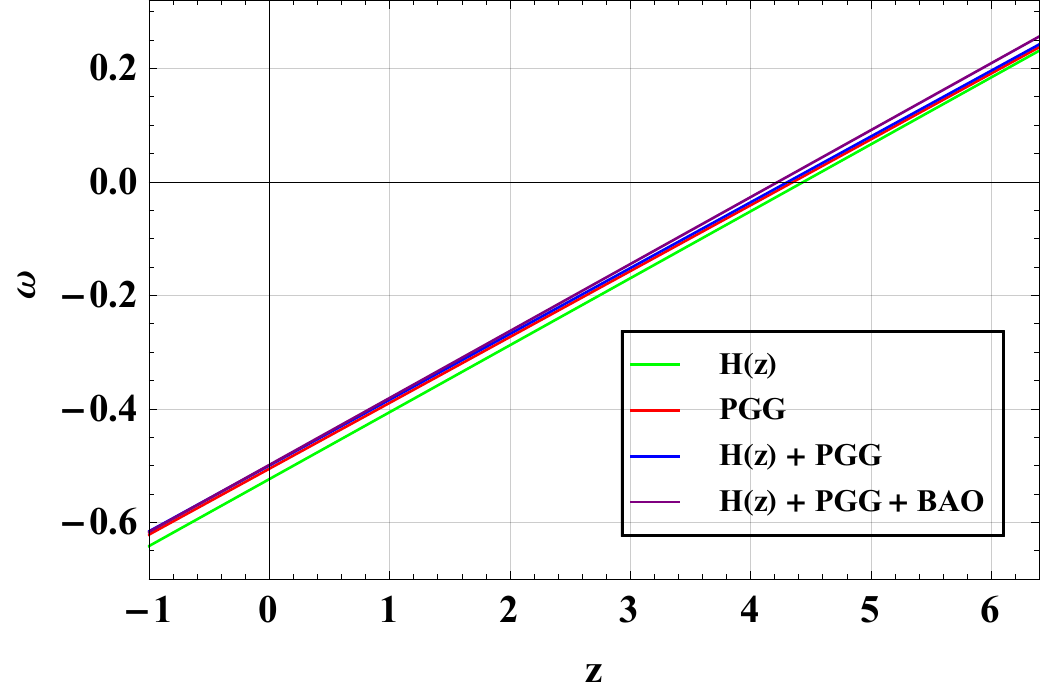}}\par
\caption{ {\it The evolution of the physical parameters $ q,~\rho,~p, and ~\omega $. }}
\label{qz2}
\end{figure}

\section{ Cosmographic analysis of the model}\label{Fourth section}
We analyze the physical parameters, including the deceleration parameter $ q $, energy density $ \rho $, isotropic pressure $ p $, and the equation of state (EoS) parameter $ \omega $ as outlined in Eqs. (\ref{13})-(\ref{18}), taking into account the presence of the Rastall parameter ($ \lambda \neq 0 $). Utilizing the optimal values for the model parameters $ \omega_0 $, $ \omega_1 $, and $ \lambda $ from Table \ref{tab1}, we examine their evolutions in Fig. \ref{qz2}.

A pivotal metric in cosmology, the Hubble parameter $ H $, delineates the universe's expansion rate. As shown in Eq. (\ref{15}), the value of $ H $ exhibits a continuous decline, asymptotically approaching zero as $ z\to -1 $ across all specified values of $ \omega_0 $, $ \omega_1 $, and $ \lambda $. That is,
\[ \lim_{z\to -1} H(z) = 0. \]

The deceleration parameter $ q $ trajectory reveals a transition from a decelerated ($ q>0 $) to an accelerated expansion ($ q<0 $) phase, marked by a change in its sign, as illustrated in Fig. \ref{qz2}. The transition point at which the deceleration parameter $ q $ changes sign, indicating a shift from deceleration to acceleration in the universe’s expansion, occurs at $ z \simeq 3.59 $ for the Hubble dataset, $ z \simeq 3.61 $ for the PGG (Pantheon+Gold+Gamma Ray Burst) dataset, $ z \simeq 3.90 $ for the Hubble+PGG dataset, and $ z \simeq 3.73 $ for the Hubble+PGG+BAO dataset. Furthermore, Fig. \ref{qz2} illustrates the monotonous decrease of the energy density $ \rho $ over time, ultimately diminishing into insignificance in the distant future. Energy density should be finite, positive, and maximum in the early stages of the universe's evolution while decreasing steadily toward later times, as shown in Fig. \ref{qz2}. The pressure's $ p $ behavior to redshift $ z $ is illustrated in Fig. \ref{qz2} for the best-fit values of $\omega_0$, $\omega_1$, and $\lambda$ utilized in the model. It transitions from positive to negative values, maintaining negativity at the present epoch ($z=0$), and appears to approach zero in the far future ($z=-1$). Positive pressure plays a significant role in the early structure formation of the universe, while the negative values of $ p $ signify the accelerating nature of the cosmos. As $ z \to -1 $, our model assumes the feature of a dust-filled universe ($ p=0 $).

The equation of state (EoS) parameter $ \omega $ in Eq. (\ref{13}) exhibits a linear dependency on the redshift $ z $. By graphically examining $ \omega $'s behavior in Fig. \ref{qz2}, distinct phases in cosmic evolution become apparent. By definition, $ \omega $ is equal to the ratio of pressure $ p $ to energy density $ \rho $ expressed as $$ \omega = \frac{p}{\rho}. $$ $ \omega $ characterizes different cosmological models based on its value. When $ \omega $ lies between $ 0 $ and $ 1 $, the universe is described by the perfect fluid. For $ \omega = 1 $, where $ p = \rho $, the model is referred to as a stiff matter field. $ \omega = 1/3 $ signifies a radiation-dominated state. In the case of $ \omega = 0 $, we have a dust-filled universe, meaning there is no pressure, but the energy density remains non-zero. A negative EoS parameter ($ \omega < 0 $) indicates a dark energy-dominated universe. Specifically,  $ \omega = -1 $ corresponds to the $ \Lambda $CDM model. When $ \omega $ falls between $ -1 $ and $ -1/3 $, it describes a quintessence dark energy model, while $ \omega < -1 $ defines a ghost or phantom model. For our model, the universe initially behaves akin to a perfect fluid. However, as time progresses, it shifts towards a quintessence dark energy model (refer to Fig. \ref{qz2}). Notably, our model mimics an accelerated expansion characteristic of dark energy in the current epoch and indefinitely into the future.

\section{ Physical features of the model}\label{Fifth section}
\subsection{ Energy Conditions}

In spacetime theories like GR, energy conditions are fundamental criteria applied to the stress-energy tensor of matter. These conditions reflect the expectation that energy density remains positive. Energy conditions can be understood geometrically, physically, and effectively \cite{Curiel:2014zba, Singh:2022jue, Shaily:2024rjq}. The geometric approach relies on geometric tensors like the Ricci or Weyl tensor, while the physical approach directly considers the stress-energy tensor. The effective formulation, assuming the Einstein Field Equations (EFEs) and non-exotic matter, is equivalent to geometric and physical formulations. Curiel and Kontou \cite{Curiel:2014zba, Kontou:2020bta} explore various energy conditions such as the weak energy condition (WEC), null energy condition (NEC), strong energy condition (SEC), and dominant energy condition (DEC). In this context, we use expressions like $ \rho \geq 0, $ $ \rho+p \geq 0 $ for WEC, $ \rho+p \geq 0 $ for NEC, $ \rho+p \geq 0 $, $ \rho+3p \geq 0 $ for SEC, and $ \rho > |p| \geq 0 $ for DEC as the basis for discussing these conditions.

\begin{figure}
\centering 
	 { \includegraphics[scale=0.45]{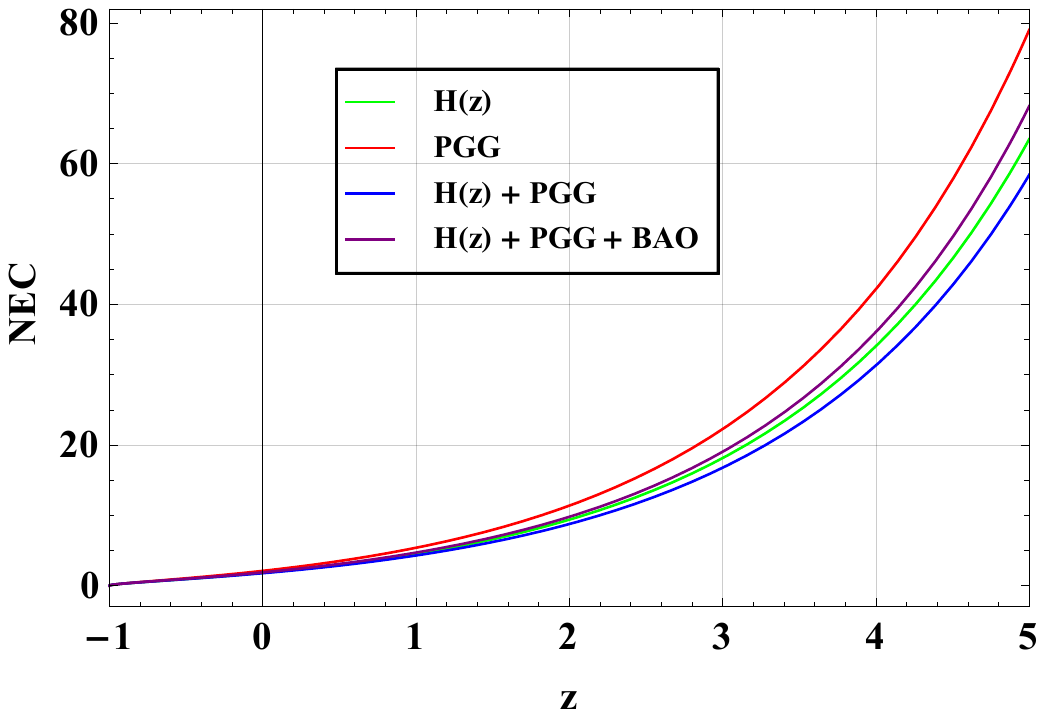}} \hfill
	 { \includegraphics[scale=0.45]{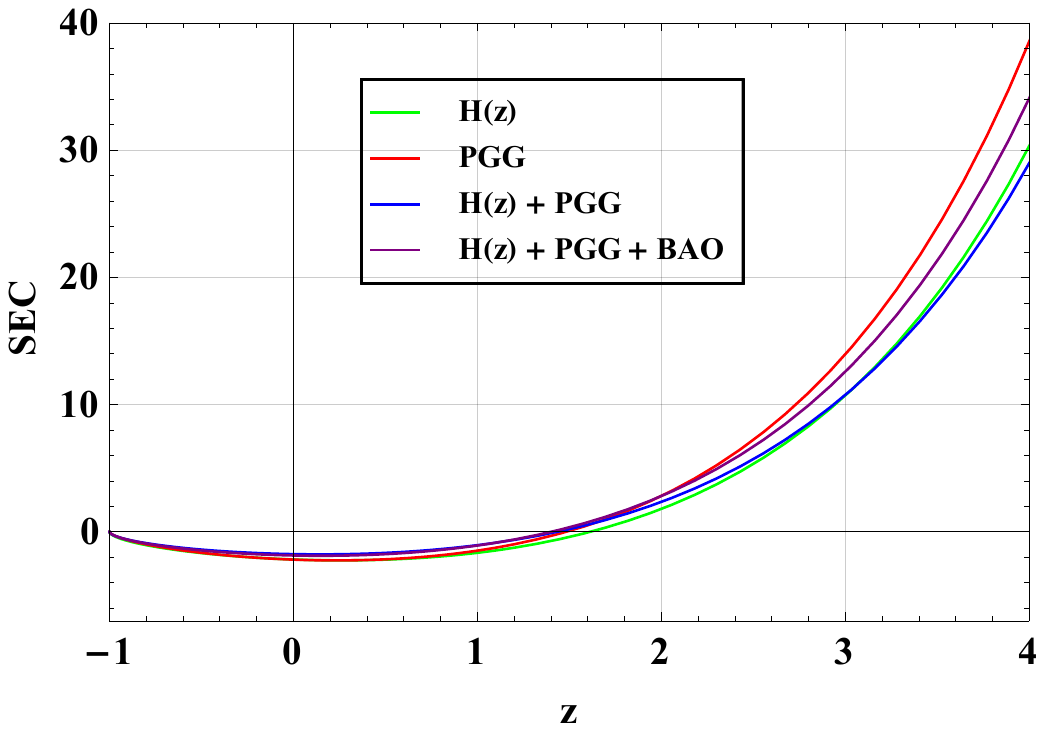}}\par 
	 { \includegraphics[scale=0.45]{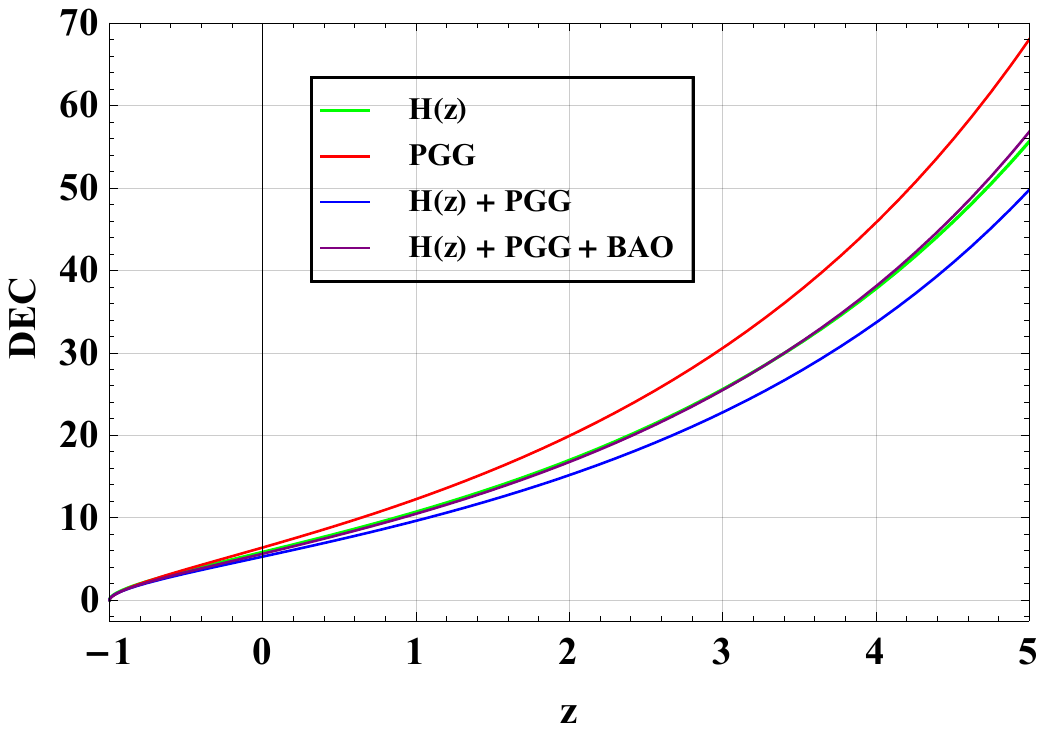}} 
\caption{ {\it The variation of the energy conditions in terms of the redshift $ z $.}}
\label{EC-z}
\end{figure}

Energy conditions are essential for understanding the universe on a large scale. These energy limits are important in explaining the existence and formation of matter. They help differentiate between ordinary and exotic matter distributions and are essential for confirming the validity of the theoretical framework. Visser's \cite{Visser:1997qk, Visser:1997tq} theoretical prediction, based on energy conditions, suggests that a negative cosmological constant might explain the violation of the SEC in the cosmological past. Notably, this violation does not necessarily mean a breach of the NEC. Even with exotic dark energy or fundamental mechanisms, violating the SEC does not automatically violate the NEC. Fig. \ref{EC-z} shows that both the NEC and the DEC hold. In contrast, the SEC is consistently violated across various best-fit values of the model parameters listed in Table \ref{tab1}. This violation implies the universe's accelerated expansion in the distant future \cite{Singh:2019fpr, Singh:2024ckh, Bolotin:2015dja, Visser:1997qk, Visser:1997tq, Singh:2023bjx}.

\subsection{ Statefinder Diagnostic}
We explored several analytical tools to investigate the geometric properties of various dark energy models, including the statefinder diagnostic method introduced by Singh et al. \cite{Singh:2015hva}. This method hinges on two crucial parameters denoted as $ r $ and $ s $. These parameters hold significant relevance, initially proposed by Sahni et al. \cite{Sahni:2002fz} and Alam et al. \cite{Alam:2003sc}. Specifically, $ r $ is linked to the jerk parameter $ j $, signifying the rate of change of the deceleration parameter $ q $, while $ s $ is intricately derived from $ r $ and $ q $, offering more profound insights into the system's dynamics. The formulae for the statefinder parameters ($ r $ and $ s $) are as follows
	\begin{equation}\label{25}
		r(z)=q+q^2+\frac{1}{(1+z)} \frac{dq}{dz},
	\end{equation}
	\begin{equation}\label{26}
		s(z)=\frac{r-1}{3(q-\frac{1}{2})},
	\end{equation}
where $ q(z) \neq \frac{1}{2} $. 
			
\begin{figure}[ht]
\centering  
	 { \includegraphics[scale=0.4]{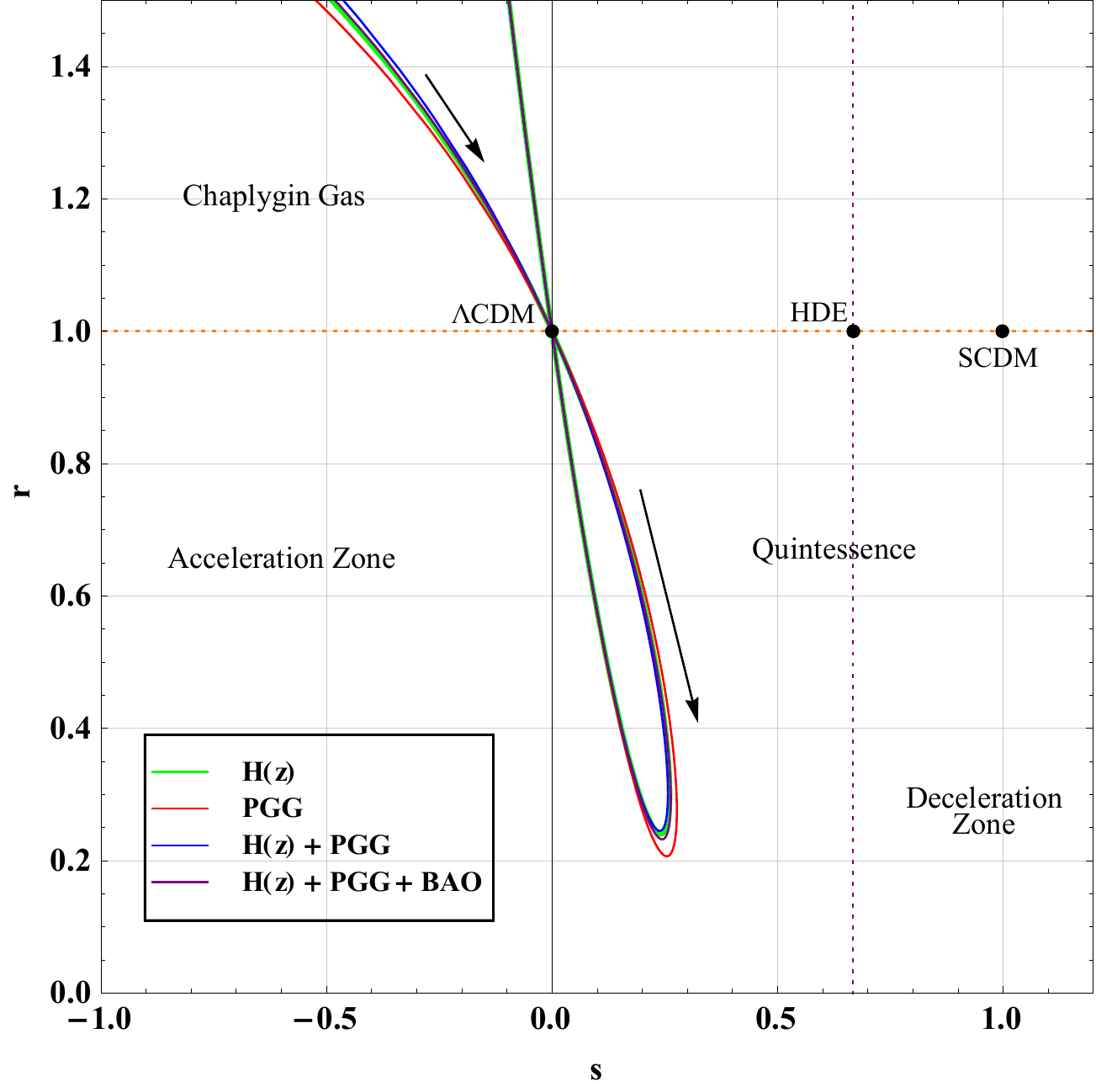}} \hfill
	 { \includegraphics[scale=0.4]{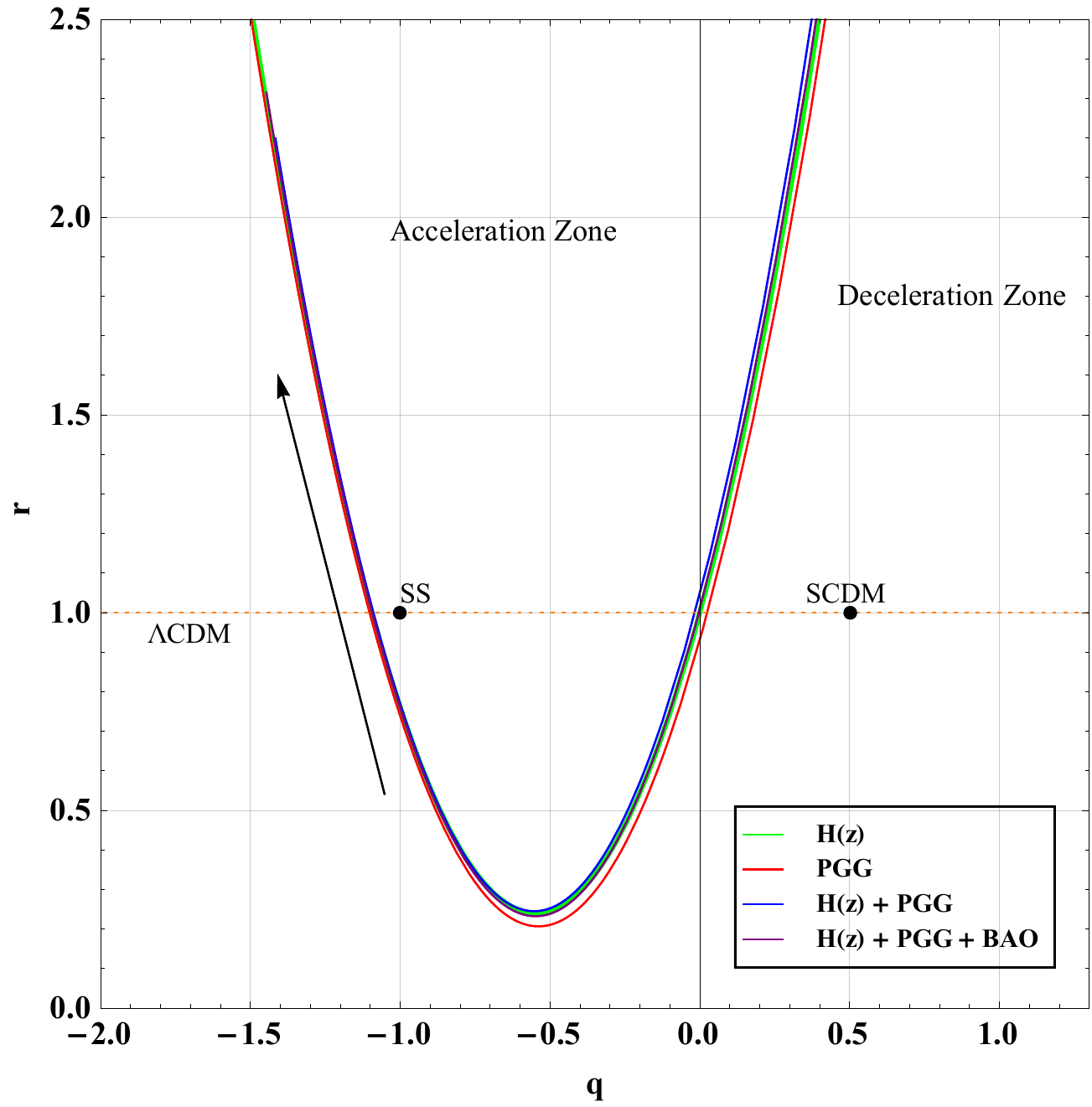}}
\caption{ {\it The Statefinder $ s-r $ and $ q-r $ plots in the Rastall gravity.}}
\label{srqr}
\end{figure}
In Fig. \ref{srqr}, the left panel showcases trajectories in the $ s-r $ plane, illustrating their traversal through the Chaplygin gas region ($ r>1, s<0 $) and quintessence region ($ r<1, s>0 $) before converging at the $ \Lambda $CDM point ($ r=1, s=0 $). On the other hand, the right panel depicts trajectories in the $ q-r $ plane, originating from the deceleration zone ($ q>0 $), diverging from the Standard Cold Dark Matter (SCDM) model ($ r=1, q=0.5 $) and Steady State model ($ r=1, q=-1 $), and subsequently entering the acceleration zone ($ q<0 $).

\subsection{ Scalar Field Analysis}

 \qquad In recent years, the quintessence model has garnered significant attention in cosmology. This model is characterized by a quintessence-like scalar field because the equation of state parameter converges in the range $ -1 < \omega < -\frac{1}{3} $ and is consistent with current observational data at later times. According to the No-Go theorem, the equation of state (EoS) parameter for a standard scalar field model, described by a Lagrangian of the form $ L = L(\phi, \partial_\mu \phi \partial^\mu \phi) $, cannot traverse the Quintom line boundary ($ \omega=-1 $) \cite{Xia:2007km, Cai:2007qw}.

To ensure compatibility with recent observational datasets, models must have an equation of state parameter close to $ \omega \simeq -1 $. This necessitates that the kinetic energy of the scalar field, $ \dot{\phi}^{2} $, be significantly smaller than the potential energy, $ V(\phi) $, i.e., $ \dot{\phi}^{2} << V(\phi) $. When $ \omega \simeq -1 $, various models can be considered to explain the universe's acceleration; moreover, these models can be adapted for inflationary purposes. In this context, our focus is on exploring cosmological models that employ scalar fields within the framework of Rastall gravity. This approach allows us to investigate potential modifications to standard cosmology and to understand the implications of alternative gravitational theories on the dynamics of scalar fields and cosmic acceleration.

 Since our model does not cross the Quintom line and the EoS parameter $ \omega $ converges in the range $ (-1,-\frac{1}{3}) $, therefore it resembles a quintessence model and not a Quintom model. 

The following action is defined in Einstein's theory of gravity
\begin{equation}\label{15a}
S=\frac{c^4}{16\pi G}\int R\sqrt{-g}d^{4}x+S_m,
\end{equation}
where $ S_m $ is the action for the quintessence-like scalar field denoted by $ {S_m}_{qu} $. Here, we take $ c=1 $ to normalize the quantity.

The action for the quintessence-like scalar field is given by
\begin{equation}\label{sf1}
	{S_m}_{qu} = \int \left( -\frac{1}{2} \partial_\mu \phi_{qu} \partial^\mu \phi_{qu} - V(\phi_{qu}) \right) \sqrt{-g} d^4x.
\end{equation}

\begin{figure}
	\centering 
	{ \includegraphics[scale=0.45]{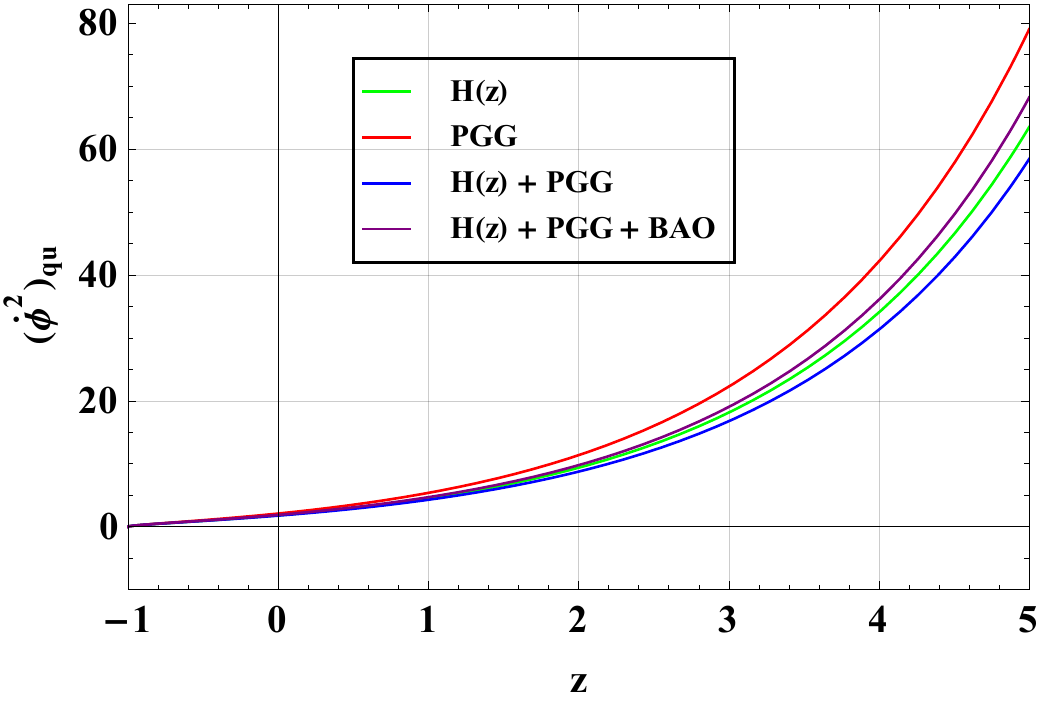}} \hfill
	{ \includegraphics[scale=0.45]{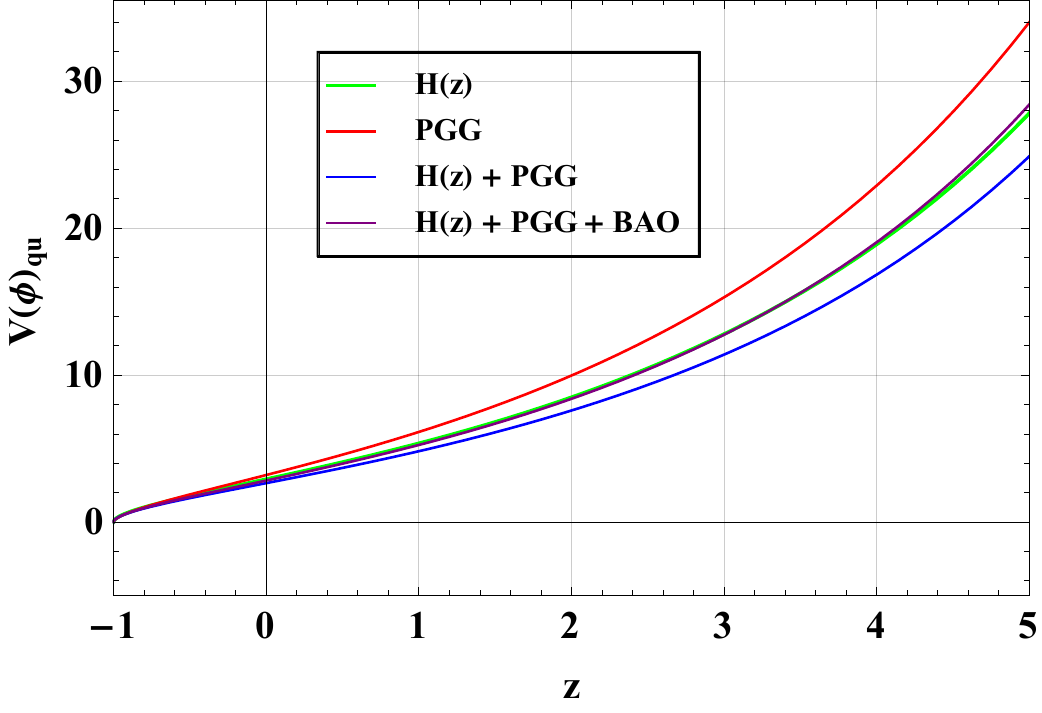}} 
	\caption{\it The evolution of the kinetic and the potential energies of the quintessence-like scalar field.}
	\label{scalar-z}
\end{figure}

In this context, the scalar potential $ V(\phi) $ corresponds to the self-interacting scalar field $ \phi $. Since $ \phi $ is time-dependent, it can be treated as a perfect fluid with energy density $ \rho_\phi $ and pressure $ p_\phi $. We consider energy density $ {\rho_\phi}_{qu} $ and pressure $ {p_\phi}_{qu} $ associated with the quintessence-like scalar field in the framework of FLRW cosmology as
\begin{equation}\label{sf3}
	{\rho_\phi}_{qu} = \frac{1}{2}{{\dot{\phi}}^2}_{qu} + V(\phi_{qu}), ~~ {p_\phi}_{qu} = \frac{1}{2}{{\dot{\phi}}^2}_{qu} - V(\phi_{qu}),
\end{equation}

The kinetic energy $ \frac{1}{2}{{\dot{\phi}}^2}_{qu} $ along with potential energy $ V(\phi_{qu}) $ corresponding to the quintessence-like scalar field are given by

\begin{equation}\label{sf5}
	\frac{1}{2}{{\dot{\phi}}^2}_{qu} = \frac{3 e^{2B} \left(-1+4\lambda \right) \left(1+\omega_0+\omega_1 z\right)}{2\left( -1+3\lambda \left(1+\omega_0+\omega_1 z\right) \right)},
\end{equation}

\begin{equation}\label{sf7}
	V(\phi_{qu}) =  -\frac{3 e^{2B} \left(-1+4\lambda \right) \left(-1+\omega_0+\omega_1 z\right)}{-2+6\lambda \left(1+\omega_0+\omega_1 z\right)}.
\end{equation}

In Fig. \ref{scalar-z}, the plot illustrates the evolution of various energy components—specifically, the kinetic energy term $ \frac{1}{2}{\dot{\phi}^2}_{qu} $ and the potential energy term $ V(\phi_{qu}) $—for the quintessence-like scalar field as functions of the redshift $ z $ within the framework of Rastall gravity, encompassing all the observational datasets. 

Initially, at the onset of the universe's evolution, the values for the kinetic energy of the quintessence-like field and the potential energy of the quintessence-like field are infinitely large. As time progresses, these values decrease monotonically, asymptotically approaching zero as $ z \to -1 $. Among these, the quintessence-like field's kinetic energy diminishes more rapidly than its potential energy. Both the potential and kinetic energies of the quintessence-like field exhibit similar diminishing behavior over time. 

\section{ Perturbation Analysis}\label{Perturbations section}
The stability of astrophysical models is a crucial aspect, and it can be evaluated by analyzing how well they withstand disturbances. In the context of cosmic evolution, even the smallest perturbations in the density of a homogeneous and isotropic fluid significantly influence the universe's large-scale structure formation \cite{Tsagas:2002sd}. These perturbations occur because the pressure is relatively weak compared to the overwhelming gravitational force. As a result, the massive cosmic structures we observe today, such as galaxies and galaxy clusters, originated from these initial perturbations. To assess the stability of proposed models, researchers can use various factors. We focus on the realm of linear perturbations to evaluate the stability of our theoretical solution within the framework of Rastall gravity. The fluctuations in both the Hubble parameter and energy density are given by \cite{delaCruz-Dombriz:2011oii, Sharif:2012ce, Bhardwaj:2022lrm, Jaybhaye:2022gxq, Narawade:2022jeg}
\begin{equation}\label{27}
	H(t) = H_0(t)(1+\delta_H(t)),
\end{equation}
and
\begin{equation}\label{28}
	\rho(t) = \rho_0(t)(1+\delta_\rho(t)).
\end{equation}
Here, $ H_0(t) $, $ \rho_0(t) $, $ \delta_H(t) $, and $ \delta_\rho(t) $ correspond to the baseline Hubble parameter, baseline energy density, and perturbation parameters for the Hubble parameter and energy density, respectively. 

In the realm of matter fields, we encounter first-order perturbation equations. The equation specifically used in this model is
\begin{equation}\label{29}
	\dot{\delta_\rho}(t) + 3H_0(t)\delta_H(t) = 0.
\end{equation}
\begin{figure}[ht]
\centering  
	 { \includegraphics[scale=0.5]{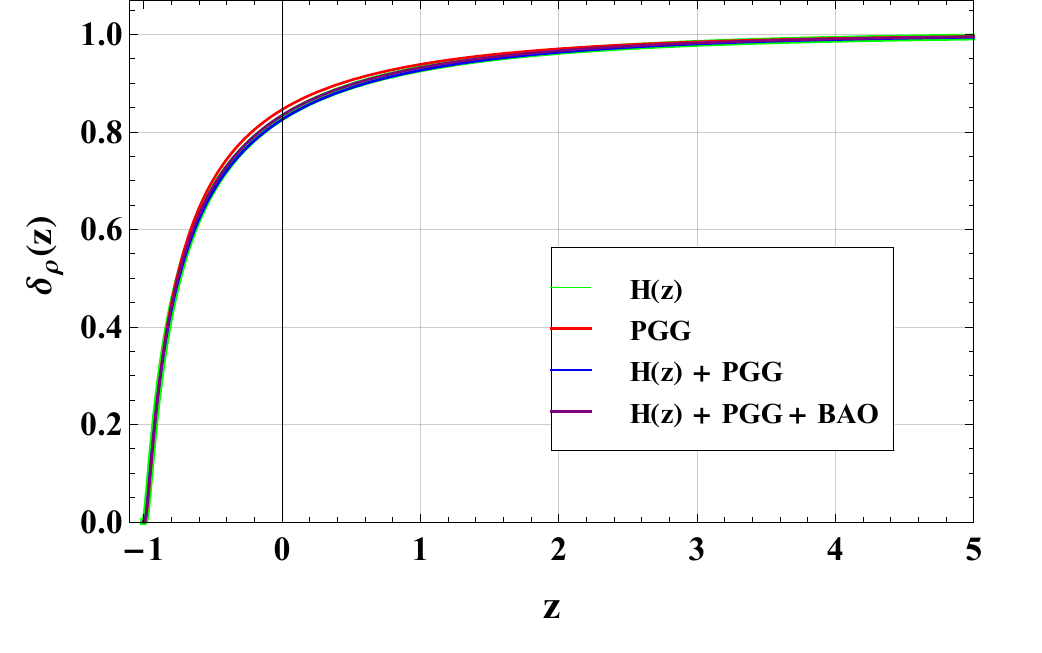}} \hfill
	 { \includegraphics[scale=0.5]{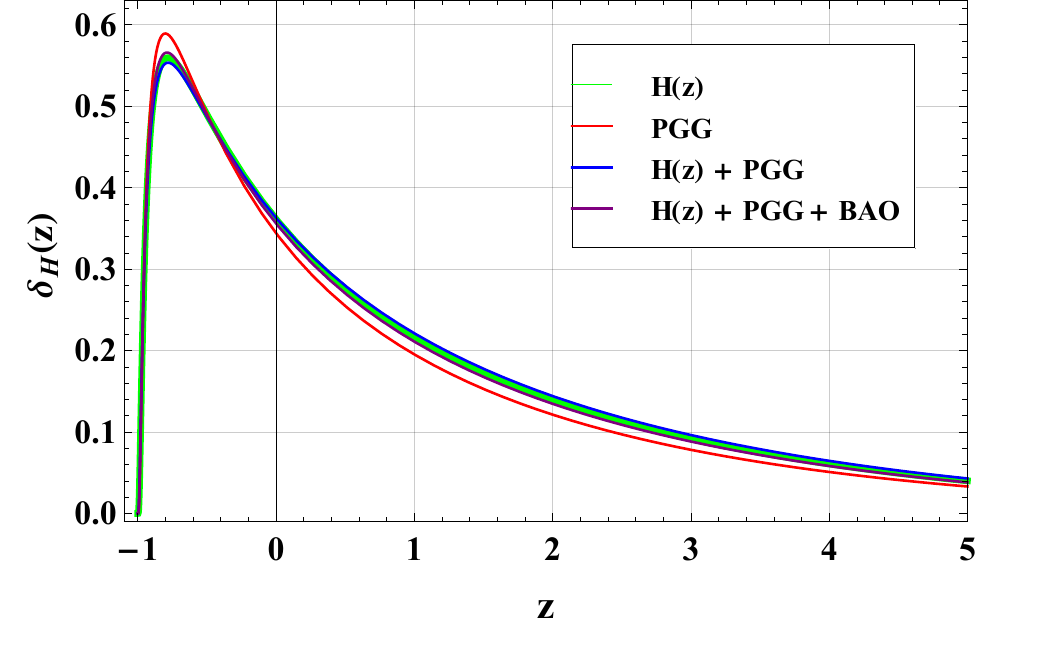}}
\caption{ {\it The illustration of the perturbations in the energy density $ \delta_\rho(z) $ and in the Hubble parameter $ \delta_H(z) $.}} 
\label{pert}
\end{figure} 
The analytical relationship between matter and geometric perturbations is established as
\begin{equation}\label{30}
b_m \delta_\rho(t) = -6\left[{H_0(t)}\right]^2\delta_H(t),
\end{equation}
where $ b_m = K\rho_{m0} $. This implies that matter perturbations determine the overall perturbation around a cosmological solution (or vice versa) concerning the GR. Now, the first-order perturbation equation used for this model can be rewritten after eliminating $ \delta_H(t) $ with the help of Eqs. (\ref{29}) and (\ref{30}) as
\begin{equation}\label{31}
\dot{\delta_\rho}(t)-\frac{b_m}{2H_0(t)}\delta_\rho(t) = 0.
\end{equation}
Integrating Eq. (\ref{31}), we get
\begin{equation}\label{32}
\delta_\rho(t) = D~\exp\left[\frac{1}{2} \int \frac{b_m}{H_0(t)} dt\right].
\end{equation}
The evolution of the Hubble perturbation $ \delta_H(t) $ can be described as
\begin{equation}\label{33}
\delta_H(t) = -\frac{b_m}{6\left[{H_0(t)}\right]^2} D~\exp\left[\frac{1}{2} \int \frac{b_m}{H_0(t)} dt\right].
\end{equation} 
Furthermore, we express the functions $ \delta_\rho(t) $ and $ \delta_H(t) $ in terms of redshift $ z $ as
\begin{equation}\label{34}
\delta_\rho(z) = D~\exp\left[\frac{1}{2} \int \frac{b_m}{(1+z)~\left[{H_0(z)}\right]^2} dz\right]
\end{equation}
and
\begin{equation}\label{35}
\delta_H(z) = -\frac{b_m}{6\left[{H_0(z)}\right]^2} D~\exp\left[\frac{1}{2} \int \frac{b_m}{(1+z)~\left[{H_0(z)}\right]^2} dz\right].
\end{equation}
Here, $ D $ represents an arbitrary integration constant.
	
The graphs in Fig. \ref{pert} are plotted using Eq. (\ref{15}) in Eqs. (\ref{34}) and (\ref{35}). The variation of perturbation parameters starts with very small values and decreases in late times to attain the value zero as redshift $ z $ approaches $ -1 $, which shows that our model is stable in the late time. Moreover, if we analyze the stability of the solution under general non-homogeneous perturbations, the scalar perturbations may exhibit instability \cite{Khlopov:1985fch, Bogdanos:2009uj, Tsujikawa:2005ju}. However, due to the complexity of the mathematical calculations involved in our model, this approach is beyond the scope of our work.

\section{ Conclusions}\label{Conclusions section}  

 We have explored the universe's evolution within the framework of the quintessence model in Rastall gravity. Our study is grounded in a variational principle, which is the basis for formulating Rastall's gravity theory. We established our model by employing a linear parametrization of the EoS parameter ($ \omega $) within the FLRW background. We derived exact solutions to the Einstein field equations in Rastall gravity.

Our analysis reveals several significant findings: The two-dimensional marginalized distributions for the model parameters $ \omega_0 $, $ \omega_1 $, and $ \lambda $ at $ 68\% $ and $ 95\% $ confidence levels were obtained, providing constraints on these parameters. The error bar plots for the Hubble parameter $ H(z) $, and the distance modulus $ \mu(z) $ across varying redshifts $ z $ demonstrated the deviations of our model from the standard $ \Lambda $CDM model in different observational datasets. The value of the Hubble parameter is compared with the $\Lambda $CDM model and some other stable models in Table \ref{tab2}.

\begin{table}
\caption Summary of the best-fit values for the Hubble parameter $ H_0 $ of our model and the $\Lambda$CDM in Rastall gravity in comparison with the exponential $ F(R) $ gravity with logarithmic corrections, the standard exponential $ F(R) $ model, the $ F(R) $ model, and the $\Lambda$CDM model of Odintsov et al. \cite{Odintsov:2023cli, Odintsov:2024lid}.
\begin{center}
\label{tab2}
\begin{tabular}{l l l l c c} 
\hline\hline
     \\ 
  {\bf Model}  &  ~~~~~{\bf Datasets}~~~~~~~~~~~~~~~~& ~~~~ $ \boldsymbol{H_0} $
        \\
        \\
        \hline
        \\  
   Rastall Gravity   &  ~~~~~$ H(z) $+Pantheon+Gold+Gamma Ray Burst+BAO &~~~ $70.8749_{-0.0001}^{+0.0002}$ 
      \\
       \\
           
  $\Lambda$CDM & ~~~~~$ H(z) $+Pantheon+Gold+Gamma Ray Burst+BAO ~~~~ &~~~ $68.9989_{-0.0098}^{+0.0081}$  
       \\
       \\
         \hline
        \\  
   Exp $+\log F(R) $ & ~~~~~CC H(z)+SNeIa+CMB+BAO ~~~~ &~~~ $ 68.92^{+1.63}_{-1.72} $  \cite{Odintsov:2024lid}
       \\
       \\
   Exp~$ +\log F(R)~ + $~axion  & ~~~~~CC H(z)+SNeIa+CMB+BAO ~~~~ &~~~ $ 69.0^{+1.72}_{-1.71} $  \cite{ Odintsov:2024lid}
  \\
  \\
  Exp $ F(R)  $ & ~~~~~CC H(z)+SNeIa+CMB+BAO ~~~~ &~~~ $ 68.84^{+1.75}_{-1.64} $  \cite{Odintsov:2024lid}
  \\
  \\
        \hline
        \\  
  $  F(R)~+ $ EDE   & ~~~~~CC H(z)+SNeIa+CMB+BAO ~~~~ &~~~ $ 68.93^{+1.61}_{-1.57} $ \cite{Odintsov:2023cli}
  \\
    \\
 $\Lambda$CDM ~~~~     & ~~~~~CC H(z)+SNeIa+CMB+BAO  & ~~~~$ 68.98^{+1.58}_{-1.60} $ \cite{Odintsov:2023cli, Odintsov:2024lid} 
       \\
       \\
\hline\hline  
\end{tabular}   
\end{center}
\end{table}

In our model, the Hubble parameter $ H $ approaches zero as $ z \to -1 $. The deceleration parameter $ q(z) $ transitions from deceleration to acceleration, aligning with current cosmological observations. The NEC and DEC are satisfied, while the SEC is violated for all optimal parameter values. The combined satisfaction of NEC and violation of SEC indicates the presence of a quintessence dark energy model, further indicating an accelerated universe expansion in the distant future. Our model exhibits intriguing behavior, transitioning from a perfect fluid model in the early universe to a quintessence dark energy model in the present epoch and late times (refer to Fig. \ref{qz2}).

 Trajectories in the $ s-r $ plane show a consistent evolution pattern and converge at the $ \Lambda $CDM point. The $ q-r $ trajectories transit from deceleration to the acceleration zone and deviate from Standard Cold Dark Matter (SCDM) and Steady State (SS) models. The variations of quintessence-like kinetic and potential energies decrease monotonically with respect to the redshift $ z $, approaching zero as $ z\to -1 $, with quintessence-like kinetic energy decreasing more rapidly. The perturbation parameters also reduce to zero as redshift approaches $ -1 $, indicating the stability of our model in late times.

In conclusion, our study demonstrates that this quintessence model in Rastall gravity provides a viable framework for understanding the universe's evolution. It successfully incorporates key observational data and offers insights into the energy conditions and stability of the universe. These findings contribute to a broader understanding of cosmology and the potential role of modified gravity theories in explaining the universe's accelerated expansion and overall dynamics. Future work will involve further investigation into the implications of these results and the potential for observational verification.

\section*{Acknowledgement} The author J. K. Singh thanks the Dept. of Mathematics, NSUT, New Delhi-78,  India for providing the necessary facilities where a part of this paper has been completed. The authors express their sincere gratitude to the respected referees for their valuable comments and suggestions.

\end{document}